\documentclass[aps]{revtex4}
\usepackage[colorlinks=true, pdfstartview=FitV, linkcolor=blue, citecolor=red, urlcolor=magenta, breaklinks=true]{hyperref}
\usepackage{graphicx}  
\usepackage{amsfonts}
\begin{document}
\title{Quantum-corrected finite entropy of noncommutative acoustic black holes}
\author{M. A. Anacleto}
\email{anacleto@df.ufcg.edu.br}
\author{F. A. Brito}
\email{fabrito@df.ufcg.edu.br}
\author{G. C. Luna}
\email{gabrielacluna@hotmail.com}
\author{E. Passos}
\email{passos@df.ufcg.edu.br}

\affiliation{Departamento de F\'{\i}sica, Universidade Federal de Campina Grande, Caixa Postal 10071, 58109-970 Campina Grande, Para\'{\i}ba, Brazil}
\author{J. Spinelly}
\email{jspinelly@uepb.edu.br}
\affiliation{Departamento de F\'isica-CCT, Universidade Estadual da Para\'iba
\\
Juv\^encio Arruda S/N, Campina Grande, PB, Brazil}
\begin{abstract} 

In this paper we consider the generalized uncertainty principle in the
tunneling formalism via Hamilton-Jacobi method to determine the quantum-corrected Hawking temperature and entropy
for 2+1-dimensional noncommutative acoustic black holes. In our results  we obtain an area entropy, a correction logarithmic in leading order, a correction term in subleading order proportional to the radiation temperature  associated with the  noncommutative acoustic black holes and an extra term that depends on a conserved charge. 
Thus, as in the gravitational case, there is no need to introduce the ultraviolet cut-off  and divergences are eliminated.

\end{abstract}
\maketitle
\pretolerance10000

\section{Introduction}
The study of acoustic black holes was proposed in 1981 by Unruh~\cite{Unruh} and has been extensively studied in the literature~\cite{MV, Volovik, others}. Acoustic black holes was found to possess many of the fundamental properties of black holes in general relativity and has been developed to investigate the Hawking radiation and other phenomena for understanding quantum gravity. Thus, many fluid systems have been investigated on a variety of analog models of acoustic black holes, including gravity wave~\cite{RS}, water~\cite{Mathis}, slow light~\cite{UL}, optical fiber~\cite{Philbin} and  electromagnetic waveguide~\cite{RSch}. The models of superfluid helium II~\cite{Novello}, atomic Bose-Einstein condensates~\cite{Garay,OL} and one-dimensional Fermi degenerate noninteracting gas~\cite{SG} have been proposed to create an acoustic black hole geometry in
the laboratory. A relativistic version of acoustic black holes has been presented in~\cite{Xian,ABP}.

In Ref.~\cite{Rinaldi} was investigated (1 + 1)-dimensional acoustic black hole entropy by the brick-wall method. In order to obtain a finite result, they had to introduce the ultraviolet cut-off. So their calculation suggested that analog black hole entropy has the “cut-off problem” similar to that of gravitational black hole entropy. More recently in~\cite{Rinaldi:2011aa} the author uses transverse modes in order to cure the divergences.

The study on the statistical origin of black hole entropy has been extensively explored by several authors --- see for instance ~\cite{Wilczek, Magan:2014dwa,Solodukhin:2011gn}. 
In Ref.~\cite{Kaul}, Kaul and Majumdar compute the lowest order
corrections to the Bekenstein-Hawking entropy. They find that the leading correction is
logarithmic, with
\[
S\sim \frac{A}{4G}-\frac{3}{2}\ln\left(\frac{A}{4G} \right)+const. +\cdots
\]
on the other hand, Carlip in Ref.~\cite{Carlip:2000nv} compute the leading logarithmic corrections to the Bekenstein-Hawking entropy and  shows that the logarithmic correction is identical to that of Kaul and Majumdar, plus corrections that depend on conserved charges, as
\[
S\sim \frac{A}{4G}-\frac{3}{2}\ln\left(\frac{A}{4G} \right) +\ln[F(Q)]+const. +\cdots
\] 
where $F(Q)$ is some function of angular momentum and other conserved charges.

The brick-wall method proposed by G. 't Hooft has been used for calculations on the black hole, promoting the understanding of the origin of black hole entropy. According to G. 't Hooft, black hole entropy is just the entropy of quantum fields outside the black hole horizon. However, when one calculates the black hole statistical entropy by this method, to avoid the divergence of states density near black hole horizon, an ultraviolet cut-off must be introduced. 
The other related idea in order to cure the divergences is to consider models in which the Heisenberg uncertainty relation is modified, for example in one dimensional space, as 
\[
\Delta x\Delta p\geq \frac{\hbar}{2}\left( 1 +\alpha^2 (\Delta p)^2 \right) ,  
\]
which shows that there exists a minimal length $ \Delta x\geq \hbar\alpha $, where $ \Delta x $ and
$ \Delta p $ are uncertainties for positon and momentum, respectively, and $ \alpha $ is a positive constant
which is independent of $ \Delta x $ and $ \Delta p $. A commutation relation for the  generalized uncertainty principle (GUP) can be written as 
$[x,p]_{GUP}=i\hbar(1+\alpha^2 p^2)  $, where $ x $ and $ p $ are the positon and the momentum operators, respectively.
Thus, using the modified Heisenberg uncertainty relation the divergence in the brick-wall model are eliminated as discussed in~\cite{Brustein}.
The statistical entropy of various black holes has also been calculated via corrected state density of the GUP~\cite{XLi}. 
Thus, the results show that near the horizon quantum state density and its statistical entropy are finite. In~\cite{KN} a relation for the corrected states density by GUP has been proposed
\[
dn=\frac{d^3 x d^3 p}{(2\pi)^3}e^{-\lambda p^2},
\]
where $p^2=p^{i}p_{i}  $, and $ \lambda $ plays the role of the Planck scale and in a fluid at high energy regimes.

Recently, the authors in~\cite{ABPS} using a new equation of state density due to GUP~\cite{Zhao}, the statistical entropy of a 2+1-dimensional rotating acoustic black hole has been analyzed. It was shown that  using the quantum statistical method the entropy of the rotating acoustic black hole was calculated, and the Bekenstein-Hawking area entropy of acoustic black hole and its correction term was obtained. Therefore, considering the effect due to GUP on the equation of state density,  no cut-off is needed \cite{Zhang} and the divergence in the brick-wall model disappears. 

There are several approaches to obtain the Hawking radiation and the entropy for black-holes. One of them is the 
Hamilton-Jacobi method which is based on the work of Padmanabhan and collaborators~\cite{SP} and also the effects of the self-gravitation of the particle are discarded.
In this way, the method uses the WKB approximation in the tunneling formalism for the computation of the imaginary part of the action. In~\cite{Parikh} the authors Parikh and Wilczek using the method of radial null geodesic determined the Hawking temperature and in~\cite{Jiang} this method was used by authors for calculating the Hawking temperature for different spacetimes. In Ref.~\cite{Banerjee} has been analyzed Hawking radiation considering  self-gravitation and back reaction effects in tunneling formalism. It has also been investigated in~\cite{Silva12} the back reaction effects for self-dual black hole using the tunneling formalism by Hamilton-Jacobi method. 
In~\cite{Majumder:afa} has been studied the effects of the GUP in the tunneling formalism for Hawking radiation to evaluate the quantum-corrected Hawking temperature and entropy for a Schwarzschild black hole.
Moreover, the authors in~\cite{Becar:2010zza} have discussed the Hawking radiation for acoustic black hole using  tunneling formalism.

In this paper, inspired by all of these previous work we apply the acoustic black hole metrics obtained from a relativistic fluid in a noncommutative spacetime~\cite{ABP12} via the Seiberg-Witten map to study the entropy of the acoustic black hole.
Whereas, on one hand our objective is to see if using the GUP in the tunneling formalism via Hamilton-Jacobi method  divergences are eliminated as in the gravitational case.

This is also motivated by the fact that in high energy physics both strong spacetime noncommutativity and quark gluon plasma (QGP) may take place together. Thus, it seems to be natural to look for acoustic black holes in a QGP fluid with spacetime noncommutativity  in this regime. Acoustic phenomena in QGP matter can be seen in Ref.~\cite{shk} and acoustic black holes in a plasma fluid can be found in Ref.~\cite{BH-plasma}.

Differently of the most cases studied, we consider the acoustic black hole metrics obtained from a relativistic fluid in a noncommutative spacetime. The effects of this set up is such that the fluctuations of the fluids are also affected.  The sound waves inherit spacetime noncommutativity of the fluid and may lose the Lorentz invariance. As a consequence the Hawking temperature is directly affected by the spacetime noncommutativity. Analogously to Lorentz-violating gravitational black holes \cite{syb,adam}, the effective Hawking temperature of the noncommutativity  acoustic black holes 
now is {\it not} universal for all species of particles. It depends on the maximal attainable velocity of this species.
Furthermore, the acoustic black hole metric can be identified with an acoustic Kerr-like black hole. 
It was found in~\cite{ABP12} that  the spacetime noncommutativity  affects the rate of loss of mass of the black hole.  
Thus for suitable values of the spacetime noncommutativity parameter a wider or narrower spectrum of particle 
wave function can be scattered with increased amplitude by the acoustic black hole. 
This increases or decreases the superressonance phenomenon previously studied in \cite{Basak:2002aw,SBP}.

In our study we shall focus on the Hamilton-Jacobi method to determine the entropy of an acoustic black hole using the GUP and considering the WKB approximation in the tunneling formalism to calculate the imaginary part of the action in order to determine the Hawking temperature and entropy for noncommutative acoustic black holes. We anticipate that we have obtained the Bekenstein-Hawking entropy of acoustic black holes and its quantum corrections. There is no need to introduce the ultraviolet cut-off  and divergences are eliminated.

The paper is organized as follows. In Sec.~\ref{II} we  briefly review the steps to find the noncommutative acoustic black hole metrics. In Sec.~\ref{HJ-method} we consider the WKB approximation in the tunneling formalism and apply the Hamilton-Jacobi method to compute the temperature and entropy. In Sec.~\ref{stat-ent} we compute the statistical entropy. We develop explicitly computations for the magnetic and electric 
sectors. Finally in Sec.~\ref{conclu} we present our final considerations.

\section{noncommutative acoustic black hole}
\label{II}
In this section let us briefly review the steps to find the noncommutative acoustic black hole metric in (3+1) dimensions from quantum field theory.
The noncommutativity is introduced by modifying its scalar and gauge sector by replacing the usual product of fields by the Moyal product \cite{SW,SGhosh,rivelles,revnc} --- see also \cite{Cai:2007xr,Cai:2014hja} for related issues.
Thus, the Lagrangian of the noncommutative Abelian Higgs model in flat space is
\begin{eqnarray}
\label{eqAHM}
\hat{\cal L}&=&-\frac{1}{4}\hat{F}_{\mu\nu}\ast\hat{F}^{\mu\nu} 
+(D_{\mu}\hat{\phi})^{\dagger}\ast D^{\mu}\hat{\phi}+ m^2\hat{\phi}^{\dagger}\ast\hat{\phi}-b\hat{\phi}^{\dagger}\ast\hat{\phi}\ast\hat{\phi}^{\dagger}\ast\hat{\phi},
\end{eqnarray}
where the hat indicates that the variable is noncommutative and the $ \ast $-product is the so-called Moyal-Weyl product or star product which is defined in terms of a real antisymmetric matrix $ \theta^{\mu\nu}$ that
parameterizes the noncommutativity of Minkowski spacetime
\begin{eqnarray}
[x^{\mu},x^{\nu}]=i\theta^{\mu\nu}, \quad \mu,\nu=0,1,\cdots,D-1.
\end{eqnarray}
The $ \ast $-product for two fields $f(x)$ and $g(x)$ is given by
\begin{eqnarray}
f(x)\ast g(x)=\exp\left(\frac{i}{2}\theta^{\mu\nu}\partial^{x}_{\mu}\partial^{y}_{\nu}\right)f(x)g(y)\vert_{x=y}.
\end{eqnarray}
As one knows the parameter $\theta^{\alpha\beta}$ is a constant, real-valued antisymmetric $D\times D$- matrix in $D$-dimensional spacetime with dimensions of length squared. For a review see \cite{revnc}.

Now using the Seiberg-Witten map~\cite{SW}, up to the lowest order in the spacetime noncommutative parameter $\theta^{\mu\nu}$, we find the corresponding theory in a commutative spacetime in (3+1) dimensions \cite{SGhosh}
\begin{eqnarray}
\label{acao}
\hat{\cal L}&=&-\frac{1}{4}F_{\mu\nu}F^{\mu\nu}\left(1+\frac{1}{2}\theta^{\alpha\beta}F_{\alpha\beta}\right) 
+\left(1-\frac{1}{4}\theta^{\alpha\beta}F_{\alpha\beta}\right)\left(|D_{\mu}\phi|^2+ m^2|\phi|^2-b|\phi|^4\right)
\nonumber\\
&+&\frac{1}{2}\theta^{\alpha\beta}F_{\alpha\mu}\left[(D_{\beta}\phi)^{\dagger}D^{\mu}\phi+(D^{\mu}\phi)^{\dagger}D_{\beta}\phi \right],
\end{eqnarray}
where $F_{\mu\nu}=\partial_{\mu}A_{\nu}-\partial_{\nu}A_{\mu}$ and  $D_{\mu}\phi=\partial_{\mu}\phi - ieA_{\mu}\phi$.

Now, if we decompose the scalar field as  $\phi = \sqrt{\rho(x, t)} \exp {(iS(x, t))}$ into the original Lagrangian, we have
\begin{eqnarray}
\label{lagran}
{\cal L}&=&-\frac{1}{4}F_{\mu\nu}F^{\mu\nu}\left(1-2\vec{\theta}\cdot\vec{B}\right)
+\rho(\tilde{\theta}g^{\mu\nu}+\Theta^{\mu\nu}){\cal D}_{\mu}S{\cal D}_{\nu}S+\tilde{\theta} m^2\rho-\tilde{\theta}b\rho^2
+\frac{\rho}{\sqrt{\rho}}(\tilde{\theta}g^{\mu\nu}+\Theta^{\mu\nu})\partial_{\mu}\partial_{\nu}\sqrt{\rho},
\end{eqnarray}
where ${\cal D}_{\mu}=\partial_{\mu}-eA_{\mu}/S  $, $\tilde{\theta}=(1+\vec{\theta}\cdot\vec{B})$, $\vec{B}=\nabla\times\vec{A}$ and $\Theta^{\mu\nu}=\theta^{\alpha\mu}{F_{\alpha}}^{\nu}$. 
In our calculations we consider the case where there is no noncommutativity between space and time, that is $\theta^{0i}=0$ and use $\theta^{ij}=\varepsilon^{ijk}\theta^{k}$, $F^{i0}=E^{i}$ and $F^{ij}=\varepsilon^{ijk}B^{k}$.

Thus, linearizing the equations of motion around the background $(\rho_0,S_0)$, with $\rho=\rho_0+\rho_1$ and  $S=S_0+\psi$ we find the equation of motion for a linear acoustic disturbance $\psi$ given by a Klein-Gordon equation in a curved space
\begin{eqnarray}
\frac{1}{\sqrt{-g}}\partial_{\mu}(\sqrt{-g}g^{\mu\nu}\partial_{\nu})\psi=0,
\end{eqnarray}
where $g_{\mu\nu}=\frac{b\rho_0}{2c_s\sqrt{f}}\tilde{g}_{\mu\nu}$ just represents the acoustic metrics in (3+1) dimensions, 
being the metric components, $ \tilde{g}_{\mu\nu} $ given in the form
\begin{eqnarray}
\tilde{g}_{tt}&=&-[(1-3\vec{\theta}\cdot\vec{B})c^{2}_{s}-(1+3\vec{\theta}\cdot\vec{B})v^2
+2(\vec{\theta}\cdot\vec{v})(\vec{B}\cdot\vec{v})-(\vec{\theta}\times\vec{E})\cdot\vec{v}],
\nonumber\\
\tilde{g}_{tj}&=&-\frac{1}{2}(\vec{\theta}\times\vec{E})^{j}(c^{2}_{s}+1)-\left[2(1+2\vec{\theta}\cdot\vec{B})
-(\vec{\theta}\times\vec{E})\cdot\vec{v}\right]\frac{v^j}{2}+\frac{B^j}{2}(\vec{\theta}\cdot\vec{v})+\frac{\theta^j}{2}(\vec{B}\cdot\vec{v}),
\nonumber\\
\tilde{g}_{it}&=&-\frac{1}{2}(\vec{\theta}\times\vec{E})^{i}(c^{2}_{s}+1)-\left[2(1+2\vec{\theta}\cdot\vec{B})-(\vec{\theta}\times\vec{E})\cdot\vec{v}\right]\frac{v^i}{2}
+\frac{B^i}{2}(\vec{\theta}\cdot\vec{v})+\frac{\theta^i}{2}(\vec{B}\cdot\vec{v}),
\nonumber\\
\tilde{g}_{ij}&=&[(1+\vec{\theta}\cdot\vec{B})(1+c^2_{s})-(1+\vec{\theta}\cdot\vec{B})v^2
-(\vec{\theta}\times\vec{E})\cdot\vec{v}]\delta^{ij}+(1+\vec{\theta}\cdot\vec{B})v^{i}v^{j}.
\nonumber\\
f&=&[(1-2\vec{\theta}\cdot\vec{B})(1+c^2_{s})-(1+4\vec{\theta}\cdot\vec{B})v^2]
-3(\vec{\theta}\times\vec{E})\cdot\vec{v}+2(\vec{B}\cdot\vec{v})(\vec{\theta}\cdot\vec{v}).
\end{eqnarray}
We should comment that in our previous computation we assumed linear perturbations just in the scalar sector,  whereas the vector field $A_\mu$ remain unchanged.

In the following we shall focus on the planar noncommutative acoustic black hole metrics in (2+1) dimensions~\cite{ABP12} to address the issues of Hawking temperature and entropy of three-dimensional acoustic black hole in the tunneling formalism via Hamilton-Jacobi method considering the generalized uncertainty principle. For the sake of simplicity, we shall consider {\it two types} of a noncommutative spacetime medium by choosing first pure magnetic sector and then we shall focus on the  
pure electric sector.

\section{Tunneling formalism via Hamilton-Jacobi Method}
\label{HJ-method}

In this section we consider the WKB approximation in the tunneling formalism and apply the Hamilton-Jacobi method to calculate the imaginary part of the action in order to determine the Hawking temperature for a noncommutative acoustic black hole. In our calculations we assume that the classical action $ {\cal I} $ satisfies
the relativistic Hamilton-Jacobi equation to leading order in the energy
\begin{eqnarray}
g^{\mu\nu}\partial_{\mu}{\cal I}\partial_{\nu}{\cal I}+m^2=0,
\end{eqnarray}
where $m$ is the mass of the scalar particle. Thus, we can neglect the effects of the self-interaction of the particle. We begin our analysis with the pure magnetic sector (the case $\vec{B}\neq 0$ and $\vec{E}=0$), as we shall show below.

\subsection{ Pure Magnetic Sector - the case $\vec{B}\neq 0$ and $\vec{E}=0$}
The acoustic line element in polar coordinates on the noncommutative plane in (2+1) dimensions, up to an irrelevant position-independent factor, in the nonrelativistic limit { was obtained in \cite{ABP12} and is given by }
\begin{eqnarray}
ds^2&=&-(1+\theta_{z}B_{z})\big\{[(1-3\theta_{z}B_{z})c^{2}_s-(1+3\theta_{z}B_{z})(v^2_{r}+v^2_{\phi})]dt^2
-2(1+2\theta_{z}B_{z})(v_{r}dr+v_{\phi}rd{\phi})dt
\nonumber\\
&+&(1+\theta_{z}B_{z})(dr^2+r^2d\phi^2)\big\}.
\end{eqnarray}
{where $B_z$ is the magnitude of the magnetic field in the $z$ direction, $ \theta_z $ is the noncommutative parameter, 
$c_s=\sqrt{dh/d\rho}$ is the sound velocity in the fluid and $v$ is the fluid velocity.}
We consider the flow with the velocity potential $\psi(r,\phi) = A\ln{r} + B\phi$  whose velocity profile in polar coordinates on the plane is  given by
\begin{eqnarray}
\vec{v}=\frac{A}{r}\hat{r}+\frac{B}{r}\hat{\phi},
\end{eqnarray}
where $B$ and $A$ are the constants of circulation and draining rates of the fluid flow.

Let us now consider the transformations of the time and the azimuthal angle coordinates as follows 
\begin{eqnarray}
&&d\tau=dt+\frac{(1+2\theta_{z}B_{z})Ardr}
{[(1-3\theta_{z}B_{z})c^2_s r^2-(1+3\theta_{z}B_{z})A^2]},
\nonumber\\
&&d\varphi=d\phi+\frac{ABdr}{r[c^2_s r^2-A^2]}.
\end{eqnarray}
In these new coordinates the metric becomes
\begin{eqnarray}
\label{ELB}
ds^2=-(1-2\Theta)\left[1-\frac{(1+6\Theta)(A^2+B^2)}{c^2_sr^2}\right]d\tau^2
+\tilde{\theta}\left[1-\frac{(1+6\Theta)A^2}{c^2_s r^2}\right]^{-1}dr^2
-\frac{2{\tilde\theta}^2B}{c_s}d\varphi d\tau+\tilde{\theta}r^2d\varphi^2,
\end{eqnarray}
where $\Theta=\theta_{z}B_{z}$ and $\tilde{\theta}=1+2\Theta$. 

The radius of the ergosphere is given by $g_{00}(r_{e}) = 0$, whereas the horizon is given by the coordinate singularity $g_{rr}(r_{h}) = 0$, that is
\begin{eqnarray}
&&\tilde{r}_{e}=(1+6\Theta)^{1/2}r_e, \quad\quad r_e=\sqrt{{r}_{h}^2+\frac{B^2}{c^2_s}}, 
\\
&&\tilde{r}_{h}=(1+6\Theta)^{1/2}r_h, \quad\quad r_h=\frac{|A|}{c_s}.
\end{eqnarray}

Now for $B=0$ (no rotation), we have the metric of stationary acoustic black hole given by
\begin{eqnarray}
\label{ELBNR}
ds^2=-f(r)d\tau^2+f(r)^{-1}dr^2+\tilde{\theta}r^2d\varphi^2,
\end{eqnarray}
where $ f(r)= (1-2\Theta)\left(1-\frac{\tilde{r}^2_h}{r^2} \right)$, with $ c_s=1 $. Thus, we obtain the Hawking temperature of the noncommutative acoustic black hole in terms of $ \tilde{r}_h $ as 
\begin{eqnarray}
\tilde{T}_{h}=\frac{f^{\prime}(\tilde{r}_h)}{4\pi}=\frac{\left(1-2\Theta\right)}{2\pi \tilde{r}_{h}} + O(\Theta^2),
\end{eqnarray}
or in terms of $r_h$, we have
\begin{eqnarray}\label{Htemp}
\tilde{T}_{h}=\left(1-5\Theta\right)T_h + O(\Theta^2), 
\end{eqnarray}
where $ T_h=(2\pi {r}_{h})^{-1}$ is the Hawking temperature of the acoustic black hole.

Near the horizon, when $ r\rightarrow \tilde{r}_h $, we have $ f(r)\approx 2a(r-\tilde{r}_h)$, where $a$ is the surface gravity of acoustic black hole horizon. 

We shall now consider the Klein-Gordon equation for a scalar field $ \Phi $ and apply the tunneling method via the Hamitlon-Jacobi ansatz
\begin{eqnarray}
\left[\frac{1}{\sqrt{-g}}\partial_{\mu}\left(\sqrt{-g}g^{\mu\nu}\partial_{\nu}\right)-\frac{m^2}{\hbar^2}\right]\Phi=0
\end{eqnarray}
where $m$ is the mass of the scalar particle. In this way, using the WKB approximation, we can write
\begin{eqnarray}
\Phi=\exp \left[\frac{i}{\hbar}{\cal I}(t,r,\varphi)\right].
\end{eqnarray}
Then, to the lowest order in $\hbar$, we have
\begin{eqnarray}
\label{wkb}
g^{\mu\nu}\partial_{\mu}{\cal I}\partial_{\nu}{\cal I}+m^2=0.
\end{eqnarray}
Considering the metric (\ref{ELBNR}) the equation (\ref{wkb}) becomes
\begin{eqnarray}
-\frac{1}{f(r)}(\partial_{t}{\cal I})^2+f(r)(\partial_{r}{\cal I})^2+\frac{1}{\tilde{\theta}r^2}(\partial_{\varphi}{\cal I})^2+m^2=0.
\end{eqnarray}
Now due to the symmetries of the metric, we can suppose a solution of the form
\begin{eqnarray}
{\cal I}=-Et  + W(r) + J_{\varphi}\varphi.
\end{eqnarray}
As a consequence
\begin{eqnarray}
\partial_t {\cal I}=-E, \quad \partial_r {\cal I}=\frac{dW(r)}{dr} \quad \partial_{\varphi} {\cal I}=J_{\varphi}.
\end{eqnarray}
where $ J_{\varphi}$ is a constant and the classical action is given by
\begin{eqnarray}
{\cal I}=-Et  + \int dr\frac{\sqrt{E^2-f(r)(m^2-\frac{J^2}{\tilde{\theta}r^2}})}{f(r)}  + J_{\varphi}\varphi.
\end{eqnarray}
In this way, by taking the near-horizon approximation,
$ f(r)\approx 2a(r-\tilde{r}_h)$, the spatial part of the action function, reads
\begin{eqnarray}
W(r)=\frac{1}{2a}\int dr\frac{\sqrt{E^2-2a(r-\tilde{r}_h)(m^2-\frac{J^2}{\tilde{\theta}r^2}})}{(r-\tilde{r}_h)}
=\frac{2{\pi} i E}{2a}.
\end{eqnarray}
Therefore, the tunneling probability for a particle with energy $ E $ is given by
\begin{eqnarray}
\Gamma\simeq \exp[-2Im {\cal I}]=\frac{2{\pi} i E}{a}.
\end{eqnarray}
On the other hand, we have $ \Gamma\simeq\exp (-\beta E) $, where $ \beta $ is the inverse temperature $ \beta=1/\tilde{T}_h $. Thus, acoustic black hole temperature is given by
\begin{eqnarray}
\tilde{T}_h=\frac{a}{2\pi}.
\end{eqnarray}
Since $a=f^{\prime}(\tilde{r}_h)/2  $, we can get the temperature in terms of the horizon radius, given by
\begin{eqnarray}
\tilde{T}_h=\frac{1-2\Theta}{2\pi \tilde{r}_h}=(1-5\Theta)T_h.
\end{eqnarray}
Note that the temperature is modified due to noncommutativity of spacetime, which precisely agrees with the result (\ref{Htemp}).

\section{The statistical entropy}
\label{stat-ent}

In this section we consider the generalized uncertainty principle (GUP) in the
tunneling formalism and apply the Hamilton-Jacobi method to determine the quantum-corrected Hawking temperature and entropy for a noncommutative acoustic black hole.
Let us start with the GUP~\cite{ADV, Tawfik:2014zca}, which is an extension of \cite{KMM} 
\begin{eqnarray}
\label{gup}
\Delta x\Delta p\geq \hbar\left( 1-\frac{\alpha l_p}{\hbar} \Delta p +\frac{\alpha^2 l^2_p}{\hbar^2} (\Delta p)^2 \right),
\end{eqnarray}
where $\alpha$ is a dimensionless positive parameter, $ l_p=\sqrt{\hbar G/c^3}={M_p G}/{c^2}\approx 10^{-35}m$ is the Plank length, $ M_p =\sqrt{\hbar c/G}$ is the Plank mass and $ c $ is the velocity of light. 
Since $G$ is the Newtonian coupling constant,  the correction terms in the uncertainty relation (\ref{gup}) are due to the effects of gravity.

Now the equation (\ref{gup}) can be written as follows.
\begin{eqnarray}
\Delta p\geq \frac{\hbar(\Delta x +\alpha l_p)}{2\alpha^2 l_p^2}
\left(1- \sqrt{1-\frac{4\alpha^2 l_p^2}{(\Delta x +\alpha l_p)^2}}\right),
\end{eqnarray}
where we have chosen the negative sign. Since $ l_p/\Delta x $ is relatively small compared to unity
we can expand the equation above in Taylor series 
\begin{eqnarray}
\label{p}
\Delta p\geq \frac{1}{\Delta x}\left[1-\frac{\alpha}{2\Delta x}+ \frac{\alpha^2}{2(\Delta x)^2}+\cdots    \right].
\end{eqnarray}
As we have chosen $ G=c=k_B=1 $, so we also choose $ \hbar=1 $, and we have $ l_p=1 $. In these units 
the uncertainty principle becomes
\begin{eqnarray}
\Delta x\Delta p\geq 1 .
\end{eqnarray}
Now using the saturated form of the uncertainty principle we have
\begin{eqnarray}
E\Delta x\geq1,
\end{eqnarray}
which follows from the saturated form of the Heisenberg uncertainty principle, $ \Delta x\Delta p\geq 1 $, where $ E $ is the energy of a quantum particle. Therefore, we can rewrite equation (\ref{p}) in the form
 \begin{eqnarray}
E_G\geq E\left[1-\frac{\alpha}{2(\Delta x)}+ \frac{\alpha^2}{2(\Delta x)^2}+\cdots    \right],
\end{eqnarray}
So by using Hamilton-Jacobi method the tunneling probability of a particle with corrected energy  $E_G$ becomes
\begin{eqnarray}
\Gamma\simeq \exp[-2Im {\cal I}]=\frac{2{\pi} i E_G}{a}.
\end{eqnarray}
comparing with the Boltzmann factor ($ e^{-E/T} $), we obtain the acoustic black hole temperature 
\begin{eqnarray}
T=\tilde{T}_h\left[ 1-\frac{\alpha}{2(\Delta x)}+ \frac{\alpha^2}{2(\Delta x)^2}+\cdots   \right]^{-1}.
\end{eqnarray}
Here we will choose $ \Delta x=2\tilde{r}_h $. Thus, we have the corrected temperature due to the GUP
\begin{eqnarray}
T=\frac{(1-2\Theta)}{2\pi \tilde{r}_h }\left[ 1-\frac{\alpha}{4\tilde{r}_h }+ \frac{\alpha^2}{8\tilde{r}_h^2 }+\cdots   \right]^{-1}.
\end{eqnarray}
Here the entropy is assumed to satisfy the area law, since it depends only on the geometry of the horizon. Thus,
using the laws of black hole thermodynamics  we get the entropy in terms of horizon area of the acoustic black
hole as
\begin{eqnarray}
S&=&\int\frac{dE}{T}=\int\frac{\kappa dA}{8\pi T}=\int\frac{d\tilde{A}}{8\pi \tilde{r}_hT}
=(1+2\Theta)\int\frac{d\tilde{A}}{4}\left[ 1-\frac{\pi\alpha}{2\tilde{A}}+ \frac{\pi^2\alpha^2}{2\tilde{A}^2}+\cdots   \right],
\nonumber\\
&=&(1+2\Theta)\left[\frac{\tilde{A}}{4}-\frac{\pi\alpha}{8}\ln{\frac{\tilde{A}}{4}}-\frac{\pi^2\alpha^2}{32 \tilde{A}/4}+\cdots\right].
\end{eqnarray}
We can write this expression for entropy in terms of horizon radius as given by
\begin{eqnarray}
\label{entropy-b}
S=(1+2\Theta)\left[\frac{2\pi \tilde{r}_h}{4}-\frac{\pi\alpha}{8}\ln{\frac{2\pi \tilde{r}_h}{4}}
-\frac{\pi^2\alpha^2}{8}\tilde{T}_h+\cdots\right],
\end{eqnarray}
where $ \tilde{A}=2\pi \tilde{r}_h=(1+3\Theta)A $ is the horizon area of the noncommutative acoustic black hole and $A=2\pi {r}_h$ is the horizon area of the acoustic black hole. 
The correction terms are due to quantum effects.
The second term is a correction logarithmic  that appears in the leading order and is similar to the existing results for gravitational black hole in 4 dimensions  and that are also derived by other methods.
The third term is a correction term to the area entropy  in subleading order and is proportional to the radiation temperature, 
$ \tilde{T}_h=(1-5\Theta)T_h $, of the noncommutative acoustic black hole. The equation (\ref{entropy-b}) can be written in terms of ${r_h}$ as
\begin{eqnarray}
\label{entropy-bc}
\tilde{S}= \frac{S}{(1+2\Theta)}=(1+3\Theta)\frac{2\pi {r}_h}{4}-\frac{\pi\alpha}{8}\ln{\frac{2\pi {r}_h}{4}}
-\frac{\pi^2\alpha^2}{8}(1-5\Theta){T}_h-\frac{\pi\alpha}{8}\ln{(1+3\Theta)}+\cdots,
\end{eqnarray}
Note that the logarithmic term is obtained due to the contribution $\alpha(\Delta p)$  in the GUP, while the gravitational case, for example, for a Schwarzschild black hole the logarithmic correction is obtained from the quadratic term $ \alpha^2(\Delta p)^2 $ in the GUP.

Thus, for $ \alpha=0 $ and $ \Theta=0 $ we just reproduce the usual semiclassical area law
for the Bekenstein-Hawking entropy, $ S=A/4=2\pi r_h/4 $. 
Moreover,  choosing $\alpha=12/\pi $, we have 
\begin{eqnarray}
\label{ent-c}
\tilde{S}=(1+3\Theta)\frac{2\pi {r}_h}{4}-\frac{3}{2}\ln{\frac{2\pi {r}_h}{4}}
-18\left(1-5\Theta\right){T}_h-\frac{9}{4}\ln{[1+2\Theta + O(\Theta^2)]}+\cdots,
\end{eqnarray}
and the resulting logarithmic correction to the entropy becomes, $-3/2\ln(A/4)$ and which has the same correction for gravitational black holes in four dimensions. The last term in (\ref{ent-c}) is independent of the horizon radius and corresponds to an extra term that depends on a conserved charge $c=e(1+2\Theta)$. Notice that both the logarithmic gravitational-like term and the appearance of the conserved charge corrections are enforced by the existence of the linear correction in the GUP we are considering in Ref.~(\ref{gup}). This conserved charge is obtained from the equations of motion for the field $ A_{\mu} $ (pure magnetic sector) of Lagrangian (\ref{lagran}), ie
\begin{eqnarray}
(\nabla \times \vec{B})^j=2e\rho(1+2\Theta)u^{j}, \quad\quad j=1,2.
\end{eqnarray}
Therefore, in our model considering a linear GUP given by (\ref{gup}) (the linear correction term in the momentum uncertainty is required),
we obtain logarithmic corrections  similar to that of Kaul and Majumdar~\cite{Kaul} and one extra term that depends on a conserved charge in a manner analogous to that obtained in~\cite{Carlip:2000nv}.

It is also interesting to note that the equation (\ref{entropy-b}) has a form similar to the result presented in Ref.~\cite{Solodukhin:2011gn} and references therein,  for black holes in two dimensions that have the following correction to the entanglement entropy
\begin{eqnarray}
S=\frac{c}{6}\sigma_h + \frac{c}{6}\ln\left(\frac{\Lambda}{\epsilon} \right),
\end{eqnarray}
where $ c $ is a central charge, $\Lambda$ is an IR cut-of, $ \epsilon $ is a UV cut-of and $ \sigma_h=\psi_h/2 $ is given by the value of the field $ \psi $ on the horizon. $\psi$ is the scalar field of the Polyakov action \cite{Solodukhin:2011gn}.


\subsection{Pure Electric Sector - the case $\vec{B}=0$ and $\vec{E}\neq 0$}

In the present subsection we repeat the previous analysis for the pure electric sector ($\vec{B}=0$ and $\vec{E}\neq 0$). As in the earlier case we take the acoustic line element obtained in \cite{ABP12}, in polar coordinates on the noncommutative plane, up to first order in $\theta$, in the `non-relativistic'  limit, given by
\begin{eqnarray}
\label{me}
ds^2\!&=&\!\left[1-\frac{3\theta}{2r}\left({\cal E}_{r}A+{\cal E}_{\phi} B\right)\right]\left\{
-\left[1-\frac{(A^2+B^2+\theta r({\cal E}_{r}A+{\cal E}_{\phi}B)}{c^2_sr^2}\right]d\tau^2\right.
\nonumber\\
&+&\left.\left[1-\frac{\theta}{r}\left({\cal E}_{r}A+{\cal E}_{\phi} B\right)\right]\left[
\left(1-\frac{A^2+\theta{\cal E}_{r}Ar}{c^2_sr^2}\right)^{-1}dr^2+r^2d\varphi^2\right]
-2\left(\frac{B}{c_sr}+\frac{\theta{\cal E}_{\phi}}{2c_s}\right)rd\varphi d\tau\right\},
\end{eqnarray}
where $ \theta{\cal E}_{r}= \theta(\vec{n}\times\vec{E})_{r}$ , 
$  \theta{\cal E}_{\phi}=\theta (\vec{n}\times\vec{E})_{\phi}$. 
The radius of the ergosphere is given by $g_{00}(\tilde{r}_{e}) = 0$, whereas the horizon is given by the coordinate singularity $g_{rr}(\tilde{r}_{h}) = 0$, that is
\begin{eqnarray}\label{horizonR}
\tilde{r}_{e}=\frac{\theta{\cal E}_{r}A+\theta{\cal E}_{\phi} B}{2c^2_s}\pm\frac{1}{2}
\sqrt{\frac{(\theta{\cal E}_{r}A+\theta{\cal E}_{\phi} B)^2}{c^4_s}+4r^2_{e}}, 
\quad\quad \tilde{r}_{h_\pm}=
r_h\left[\frac{\theta{\cal E}_{r}}{2c_s}\pm \sqrt{1+\frac{(\theta{\cal E}_{r})^2}{4c^2_s}}\right]
\end{eqnarray}
where $ r_{e}=\sqrt{(A^2+B^2)/c^2_s }$  and $ r_{h}=|A|/c_s $ are the radii of the ergosphere and the horizon in the usual case. For $ \theta=0 $, we have  $\tilde{r}_{e}={r}_{e}$ and $\tilde{r}_{h}={r}_{h}  $.

Now for $B = 0$ (no rotation) and $ {\cal E}_{\phi}=0 $ with $ c_s=1 $, we have the metric of stationary acoustic black hole given by
\begin{eqnarray}
\label{mest}
ds^2= -f(r)d\tau^2+\left(1-\frac{4\theta{\cal E}_{r}r_h}{r}\right)f(r)^{-1}dr^2+\left(1-\frac{5\theta{\cal E}_{r}r_h}{2r}\right)r^2d\varphi^2,
\end{eqnarray}
where
\begin{eqnarray}
f(r)=\left(1-\frac{3\theta{\cal E}_{r}r_h}{2r}\right)\left(1-\frac{r^2_h}{r^2} -\frac{\theta{\cal E}_{r}r_h}{r}\right).
\end{eqnarray}
Now we obtain the Hawking temperature of the acoustic black hole as 
\begin{eqnarray}
T_{h}=\frac{f^{\prime}(\tilde{r}_{h+})}{2\pi}=\left(1-2\theta{\cal E}_{r}\right)\frac{1}{2\pi \tilde{r}_{h+}}+{\cal O}(\theta^2).
\end{eqnarray}
Thus, considering the tunneling formalism via Hamilton-Jacobi method with the GUP  (\ref{gup})
we obtain the corrected  temperature as
\begin{eqnarray}
T=\frac{(1-2\theta\varepsilon_r)}{2\pi \tilde{r}_{h+} }\left[ 1-\frac{\alpha}{4\tilde{r}_{h+} }+ \frac{\alpha^2}{8\tilde{r}_{h+}^2 }+\cdots   \right]^{-1}.
\end{eqnarray}

In this way, the entropy of the noncommutative acoustic black hole, near the black hole horizon, is
\begin{eqnarray}
S=(1+2\theta\varepsilon_r)\left[\frac{2\pi \tilde{r}_{h+}}{4}
-\frac{\pi\alpha}{8}\ln{\frac{2\pi \tilde{r}_{h+}}{4}}-\frac{\pi^2\alpha^2}{8}\tilde{T}_h+\cdots\right],
\end{eqnarray}
where $ 2\pi \tilde{r}_{h+} $ is the horizon area of the noncommutative acoustic black hole, ie, $ \tilde{A}=(1+\theta\varepsilon_r/2)A $ and $ A= 2\pi {r}_{h}$ the horizon area of the acoustic black hole.  
Again, the correction terms are due to quantum effects. Thus the entropy in terms of $r_h$ is
\begin{eqnarray}
\tilde{S}=\frac{S}{(1+2\theta\varepsilon_r)}=\left(1+\frac{\theta\varepsilon_r}{2}\right)\frac{2\pi {r}_{h}}{4}
-\frac{\pi\alpha}{8}\ln{\frac{2\pi {r}_{h}}{4}}-\left(1-\frac{5\theta\varepsilon_r}{2}\right)\frac{\pi^2\alpha^2}{8}{T}_h
-\frac{\pi\alpha}{8}\ln\left(1+\frac{\theta\varepsilon_r}{2}\right)+\cdots.
\end{eqnarray}
Note that the second term is a logarithmic correction that appears in the leading order as in the gravitational case in 3+1 dimensions and the third correction term is proportional to the radiation temperature of the acoustic black hole. 
 Moreover, choosing $ \alpha=12/\pi $, we have
\begin{eqnarray}
\label{entropia-e}
\tilde{S}=\left(1+\frac{\theta\varepsilon_r}{2}\right)\frac{2\pi {r}_{h}}{4}
-\frac{3}{2}\ln{\frac{2\pi {r}_{h}}{4}}-18\left(1-\frac{5\theta\varepsilon_r}{2}\right){T}_h
-\frac{3}{2}\ln\left(1+\frac{\theta\varepsilon_r}{2}\right)+\cdots.
\end{eqnarray}
The last term in (\ref{entropia-e}) is independent of the horizon radius and corresponds to an extra term 
that depends on a conserved charge $ c=e(1+\theta\varepsilon_r/2) $. This conserved charge is obtained from the equations of motion for the field $ A_{\mu} $ (pure electric sector) of Lagrangian (\ref{lagran}), i.e.,
\begin{eqnarray}
\nabla \cdot \vec{E}=2e\rho w\left(1+\frac{\theta\varepsilon_{r}}{2}\right),
\end{eqnarray}
where we have considered the normalization $ v_r\equiv u_r/w=1 $, since in the near horizon regime $v_r\sim c_s=1$.


\section{conclusions}
\label{conclu}

In summary, by considering the GUP, we derive the noncommutative acoustic black hole temperature and entropy using the Hamilton-Jacobi version of the tunneling formalism.  Moreover, in our calculations the Hamilton-Jacobi method was applied to calculate the imaginary part of the action and the GUP was introduced by the correction to the energy of a particle due to gravity near the horizon. The GUP allows us to find quantum corrections to the area law. The noncommutative nature of the spacetime also suggests the existence of corrections in terms of conserved electric charge.

\acknowledgments

We would like to thank CNPq, CAPES and PNPD/PROCAD -
CAPES for partial financial support.

\end{document}